\RequirePackage[table]{xcolor}
\documentclass{PoS}
\usepackage{setspace}
\usepackage{slashed}
\usepackage{verbatim}    
\usepackage{graphicx} 
\usepackage{wrapfig}
\usepackage{dsfont}
\usepackage{epsfig}
\usepackage{epstopdf}
\usepackage{amsmath}
\usepackage{amsfonts,amssymb}
\usepackage{cleveref}
\usepackage{xcolor}

\newcommand{\MSbar}{{\overline{\rm MS}}}

\newcommand{\be}{\begin{equation}}
\newcommand{\ee}{\end{equation}}
\newcommand{\bea}{\begin{eqnarray}}
\newcommand{\eea}{\end{eqnarray}}
\def\lsim{\mathrel{\rlap{\lower4pt\hbox{\hskip1pt$\sim$}}
    \raise1pt\hbox{$<$}}}                

\title{Four-quark operators with $\Delta$F = 2 in the GIRS scheme}
\ShortTitle{Four-quark operators with $\Delta$F = 2 in the GIRS scheme}

\author{M.~Constantinou$\,^a$, \speaker{M.~Costa}$^{, \,b,\, c, \, d}$, H.~Herodotou$^{ \,c}$, H.~Panagopoulos$^{\,c}$, G.~Spanoudes$^{\,c}$ \\
\llap{}$^a$Department of Physics, Temple University, Philadelphia, PA 19122 - 1801, USA\\
$^b$ Department of Mechanical Engineering and Material Science and Engineering, \\ Cyprus University of Technology, Limassol, CY-3036, Cyprus\\
$^c$ Department of Physics, University of Cyprus, Nicosia, CY-1678, Cyprus\\
$^d$ Rinnoco Ltd, Limassol, CY-3047, Cyprus

{\rm E-mail}: \email{marthac@temple.edu}, \email{kosta.marios@ucy.ac.cy}, \email{herodotos.herodotou@ucy.ac.cy}, \email{panagopoulos.haris@ucy.ac.cy} \email{spanoudes.gregoris@ucy.ac.cy}}

\abstract{We calculate the mixing matrices of four-quark operators that change flavor numbers by two units. Our approach employs two schemes: the coordinate-space Gauge Invariant Renormalization Scheme (GIRS) and the Modified Minimal Subtraction scheme ($\MSbar$). From our perturbative computations, we extract the conversion factors between these two renormalization schemes at the next-to-leading order. A significant challenge in the study of four-quark operators is that they mix among themselves upon renormalization. Additionally, computations in GIRS at a given order in perturbation theory require Feynman diagrams with at least one additional loop. The extraction of the conversion factors involves calculating two-point Green's functions, which include products of two four-quark operators, and three-point Green's functions, which involve one four-quark operator and two bilinear operators, with all operators located at distinct spacetime points. We investigate both parity-conserving and parity-violating four-quark operators. This calculation is relevant to the determination of Cabibbo–Kobayashi–Maskawa (CKM) matrix elements from numerical simulations using the GIRS scheme. Further details, including the GIRS anomalous dimensions obtained through the Renormalization Group (RG) equation, as well as additional results, can be found in our paper~\cite{Constantinou:2024wdb}.
\begin{center}
\includegraphics[scale=0.45]{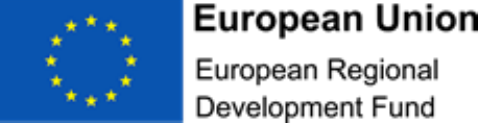}
\includegraphics[scale=0.45]{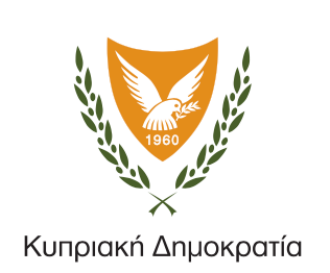}
\includegraphics[scale=0.45]{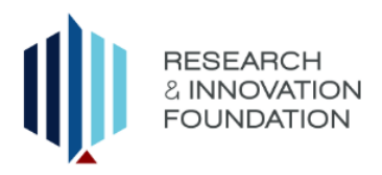}
\includegraphics[scale=0.45]{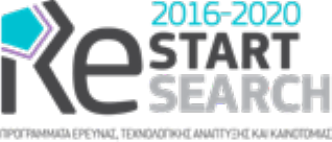}
\end{center}
}

\FullConference{The 41st International Symposium on Lattice Field Theory (LATTICE2024)\\
 28 July - 3 August 2024\\
Liverpool, UK\\}

\begin{document}
\maketitle

\section{Introduction}
The Standard Model (SM)  is a highly successful framework describing electroweak and strong interactions. Its renormalizability arises from including all relevant operators with mass dimension \( D \leq 4 \), consistent with Lorentz invariance and gauge symmetry. Higher-dimensional operators (\( D > 4 \)) are suppressed by powers of a high-energy scale \( M \), representing physics beyond the SM. This is a characteristic of effective field theory, where new physics at a scale \( M \) is integrated out, leaving higher-dimensional operators. Among these, dimension-6 four-quark operators, suppressed by \( M^{-2} \), play a key role in probing corrections to SM processes and potential new physics. Such operators offer critical insights into phenomena beyond the SM, as discussed in foundational work (see review \cite{Buchalla:1995vs}), with recent updates \cite{USQCD:2019hyg,FlavourLatticeAveragingGroupFLAG:2021npn}.

When considering lattice simulations of Quantum Chromodynamics (QCD), scalar and pseudoscalar four-quark operators inherently encapsulate weak interaction effects. Their study is especially timely given the high precision achieved in experimental measurements of CKM matrix elements and the recent experimental results from the LHCb collaboration, which have highlighted the discovery of new tetraquark states~\cite{LHCb:2022aki}. This motivates the need to explore the properties of four-quark operators numerically on the lattice, as they are central to understanding phenomena such as electroweak decays of hadrons and new physics beyond the SM. In particular, calculating the matrix elements of four-quark operators in lattice QCD can provide profound insights into these processes. Phenomenological quantities such as the so-called bag parameters are important lattice observables related to four-quark operators, with one of the most extensively studied being the $B_K$ parameter, which describes neutral $K^0 - \overline{K}^0$ meson oscillations. This parameter, along with other related bag parameters, has significant implications for hadronic decays and CP violation, making their lattice calculations crucial.

In this work, we revisit the renormalization of the four-quark operators by employing a Gauge Invariant Renormalization Scheme (GIRS)~\cite{Costa:2021iyv}, which involves Green’s functions of gauge-invariant operators in coordinate space. A similar recent study using coordinate-space renormalization prescription has been carried out in~\cite{Lin:2024mws}. GIRS is a promising renormalization prescription that does not encounter issues in lattice studies related to gauge fixing. Our goal is to provide appropriate renormalization conditions, which address the mixing of the four-quark operators and which are applicable in nonperturbative calculations on the lattice, as well as to provide the conversion factors between GIRS and the $\MSbar$ scheme. The conversion factors are regularization-independent, and thus one can compute them in dimensional regularization (DR), where perturbative computation can be performed more readily and in higher-loop order. To this end, we calculate the first quantum corrections for appropriate two-point and three-point Green’s functions in coordinate space using DR. We focus on the renormalization of four-quark operators, which are involved in flavor-changing $\Delta F=2$ processes. These are categorized into 2 sets of parity-conserving and 2 sets of parity-violating operators. More details, GIRS anomalous dimensions extracted through the Renormalization Group (RG) equation, along with additional results, can be found in our paper~\cite{Constantinou:2024wdb}.

\section{Computational Setup}

We focus on the renormalization of four-quark composite operators of the form:
\begin{equation}
{\cal O}_{\Gamma \Tilde{\Gamma}}(x) = (\bar{\psi}_{f_1}(x) \Gamma \psi_{f_3}(x)) (\bar{\psi}_{f_2}(x) \Tilde{\Gamma} \psi_{f_4}(x)),
\label{eq:fourQ}
\end{equation}
where the subscripts $f$ denote flavor indices, $\Gamma$ and $\Tilde{\Gamma}$ denote products of Dirac matrices:
\begin{eqnarray}
\Gamma, \Tilde{\Gamma} \in 1,\, \gamma_5,\, \gamma_\mu,\, \gamma_\mu \gamma_5,
  \,\sigma_{\mu\nu}, \,\gamma_5\sigma_{\mu\nu}\} \equiv
  \{S,P,V,A,T,\tilde T \}.
\end{eqnarray}
Fermion-antifermion pairs in parentheses are color singlets. 
Our primary focus is on $\Delta F = 2$ operators, which are scalar or pseudoscalar under rotational symmetry, i.e. $\Gamma = \Tilde{\Gamma}$ or $\Gamma = \Tilde{\Gamma} \gamma_5$ (with a summation intended over repeated indices). Therefore, there are five scalar and another five pseudoscalar operators.  Furthermore, there are another 10 four-quark operators ${\cal O}^F_{\Gamma \Tilde{\Gamma}}$, which, in the absence of color indices, would be linear combinations of the original operators through the Fierz–Pauli–Kofink identity (the superscript $F$ stands for Fierz):
\begin{equation}
    {\cal O}^F_{\Gamma \Tilde{\Gamma}} \equiv (\bar \psi_{f_1}\,\Gamma\,\psi_{f_4})(\bar \psi_{f_2}\,\Tilde{\Gamma}\,\psi_{f_3}). \label{O^F_XY}
\end{equation}

In order to identify which operators mix among themselves, we considered the symmetries of the QCD action: Parity $\mathcal{P}$, Charge conjugation $\mathcal{C}$, Flavor exchange symmetry $\mathcal{S} {\equiv} (f_3 \leftrightarrow f_4)$, Flavor Switching symmetries $\mathcal{S}' {\equiv} (f_1 \leftrightarrow f_3 , f_2 \leftrightarrow f_4)$ and
$\mathcal{S}'' {\equiv} (f_1 \leftrightarrow f_4 , f_3 \leftrightarrow f_2)$), with mass-degenerate quarks \cite{Frezzotti:2004wz}. Chiral symmetry can be violated in some regularizations and thus, it is not considered in the present study for identifying the mixing pattern. In particular, the operator-mixing setup which follows is also applicable in lattice regularizations that break chiral symmetry (such as Wilson fermions). Operators with the same transformation properties under these symmetries can and will mix. The parity $\mathcal{P}$ and charge conjugation $\mathcal{C}$ transformations on quarks and antiquarks are defined below:
\begin{eqnarray}
\rm{Parity:}
&&\begin{cases}
    \mathcal{P}\psi_f(x) & = \gamma_4 \ \psi_f(x_P) \\
    \mathcal{P}\bar{\psi}_f(x) &= \bar{\psi}_f(x_P) \ \gamma_4,
\end{cases} \\
\rm{Charge\,conjugation:}
&&\begin{cases}
    \mathcal{C}\psi_f(x) &= -C \ \bar{\psi}_f^T(x) \\
    \mathcal{C}\bar{\psi}_f(x) &= \psi_f^T(x) \ C,
\end{cases}
\end{eqnarray}
where $x_P = (-\textbf{x},t)$, $^{\,T}$ means transpose and the matrix $C$ satisfies: $(C \gamma_{\mu})^{T}= C \gamma_{\mu}$, $C^T=-C$ and $C^{\dagger} C=1$.

It is convenient to use a new basis where we involve sums and differences of the operators ${\cal O}_{\Gamma \Tilde{\Gamma}}$ and ${\cal O}^F_{\Gamma \Tilde{\Gamma}}$. In this way, the operators can be further decomposed into smaller, independent bases according to the discrete symmetries $P$, $S$, $CPS'$, and $CPS''$, as shown below:
\be
\begin{split}
\begin{cases}
Q_1^{S=\pm 1}\equiv \frac{1}{2}\left[{\cal O}_{VV} \pm {\cal O}^F_{VV}\right]+\frac{1}{2}\left[{\cal O}_{AA} \pm {\cal O}^F_{AA}\right]\\[0.4ex]
Q_2^{S=\pm 1}\equiv \frac{1}{2}\left[{\cal O}_{VV} \pm {\cal O}^F_{VV}\right]-\frac{1}{2}\left[{\cal O}_{AA} \pm {\cal O}^F_{AA}\right]\\[0.4ex]
Q_3^{S=\pm 1}\equiv \frac{1}{2}\left[{\cal O}_{SS} \pm {\cal O}^F_{SS}\right]-\frac{1}{2}\left[{\cal O}_{PP} \pm {\cal O}^F_{PP}\right]\\[0.4ex]
Q_4^{S=\pm 1}\equiv \frac{1}{2}\left[{\cal O}_{SS} \pm {\cal O}^F_{SS}\right]+\frac{1}{2}\left[{\cal O}_{PP} \pm {\cal O}^F_{PP}\right]\\[0.4ex]
Q_5^{S=\pm 1}\equiv \frac{1}{2}\left[{\cal O}_{TT} \pm {\cal O}^F_{TT}\right],
\end{cases}
\end{split}
\qquad \quad 
\begin{split}
&\begin{cases}
{\cal Q}_1^{S=\pm 1}\equiv \frac{1}{2}\left[{\cal O}_{VA} \pm {\cal O}^F_{VA}\right]+\frac{1}{2}\left[{\cal O}_{AV} \pm {\cal O}^F_{AV}\right],
\end{cases}\\
&\begin{cases}
{\cal Q}_2^{S=\pm 1}\equiv \frac{1}{2}\left[{\cal O}_{VA} \pm {\cal O}^F_{VA}\right]-\frac{1}{2}\left[{\cal O}_{AV} \pm {\cal O}^F_{AV}\right]\\
{\cal Q}_3^{S=\pm 1}\equiv \frac{1}{2}\left[{\cal O}_{PS} \pm {\cal O}^F_{PS}\right]-\frac{1}{2}\left[{\cal O}_{SP} \pm {\cal O}^F_{SP}\right],
\end{cases}\\
&\begin{cases}
{\cal Q}_4^{S=\pm 1}\equiv \frac{1}{2}\left[{\cal O}_{PS} \pm {\cal O}^F_{PS}\right]+\frac{1}{2}\left[{\cal O}_{SP} \pm {\cal O}^F_{SP}\right]\\
{\cal Q}_5^{S=\pm 1}\equiv \frac{1}{2}\left[{\cal O}_{T\tilde T} \pm {\cal O}^F_{T\tilde T}\right].
\end{cases}    
\end{split}
\label{Q_definitions}
\ee
The operators of Eq.~\eqref{Q_definitions} are grouped according to their mixing pattern. Therefore, the mixing matrices $Z^{S=\pm 1}$ (${\cal Z}^{S=\pm 1}$), which renormalize the Parity Conserving (Violating) operators, $Q^{S=\pm 1}_{i}$ (${\cal{Q}}^{S=\pm 1}_{i}$), take the following form:
{\small{
\be
Z^{S=\pm 1}
=
\left(\begin{array}{rrrrr}
Z_{11}\,\, & Z_{12}\,\, & Z_{13}\,\, & Z_{14}\,\, & Z_{15} \\
Z_{21}\,\, & Z_{22}\,\, & Z_{23}\,\, & Z_{24}\,\, & Z_{25} \\
Z_{31}\,\, & Z_{32}\,\, & Z_{33}\,\, & Z_{34}\,\, & Z_{35} \\
Z_{41}\,\, & Z_{42}\,\, & Z_{43}\,\, & Z_{44}\,\, & Z_{45} \\
Z_{51}\,\, & Z_{52}\,\, & Z_{53}\,\, & Z_{54}\,\, & Z_{55} 
\end{array}\right)^{S=\pm 1},
\quad
{\cal Z}^{S=\pm 1}
=
\left(\begin{array}{rrrrr}
 {\cal Z}_{11}  &0\,\,         &0\,\,         &0\,\,        &0\,\,  \\
 0\,\,         &{\cal Z}_{22}  &{\cal Z}_{23}  &0\,\,        &0\,\,  \\
 0\,\,         &{\cal Z}_{32}  &{\cal Z}_{33}  &0\,\,        &0\,\,  \\
 0\,\,         &0\,\,         &0\,\,         &{\cal Z}_{44}  &{\cal Z}_{45} \\
 0\,\,         &0\,\,         &0\,\,         &{\cal Z}_{54}  &{\cal Z}_{55}
\end{array}\right)^{S=\pm 1}.
\label{MixingMatrix}
\ee
}}

The renormalized Parity Conserving (Violating) operators,
$\hat{Q}^{S=\pm 1}$ ($\hat{\cal Q}^{S=\pm 1}$), are defined via the
equations:
\be
{\hat Q}_l^{S=\pm 1} = Z^{S=\pm 1}_{lm} \cdot Q^{S=\pm 1}_{m} ,\quad
\hat{\cal Q}^{S=\pm 1}_l = {\cal Z}^{S=\pm 1}_{lm} \cdot {\cal Q}^{S=\pm 1}_m,
\ee
where $l,m = 1,\dots ,5$ (a sum over $m$ is implied). 

Our approach entails calculating two-point and three-point Green's functions in dimensional regularization (DR) with $D = 4 - 2\epsilon$, where $\epsilon$ is the regulator. The two-point Green's functions involve products of two four-quark operators placed at nonzero distances in coordinate space. To determine all mixing coefficients, several three-point functions are required in the renormalization conditions. These three-point functions involve one four-quark operator and two bilinear operators, each placed at a different spacetime point to avoid contact singularities.

\section{One-loop Conversion Matrices between GIRS and $\MSbar$ Schemes}
\begin{figure}[ht!]
\begin{center}
  \includegraphics[scale=0.75]{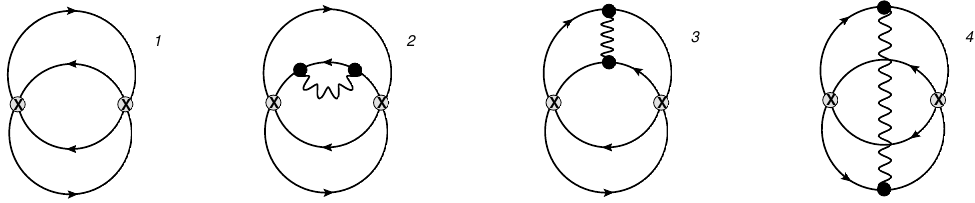} 
\caption{Feynman diagrams contributing to $G^{\rm 2pt}_{\mathcal{O}_{\Gamma \Tilde{\Gamma}};\mathcal{O}_{\Gamma' \Tilde{\Gamma'}}} (\vec{z,}z_4) \equiv \langle {\cal O}_{\Gamma \Tilde{\Gamma}} (0) \,  {\cal O}^\dagger_{\Gamma' \Tilde{\Gamma'}} (z) \rangle$, up to order $\mathcal{O}(g^2)$. Wavy (solid) lines represent gluons (quarks).}
\label{fig4Q4Q:2pt}  
\end{center}
\end{figure}

\begin{figure}[ht!]
\begin{center}
\includegraphics[scale=0.65]{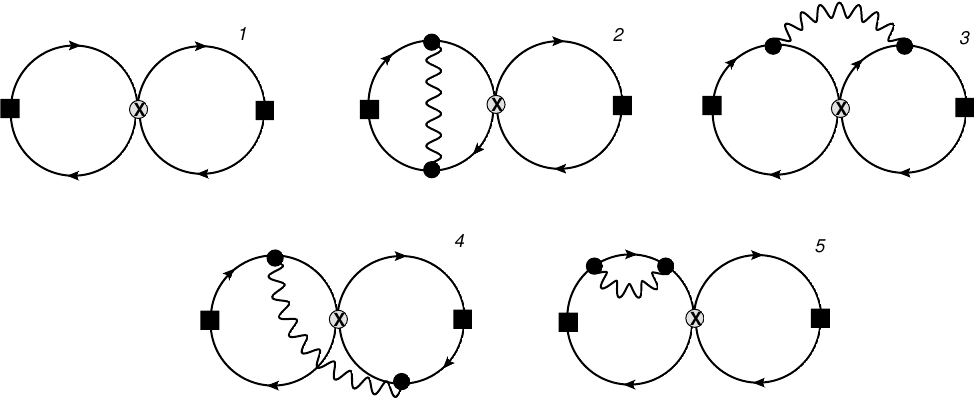} 
\caption{Feynman diagrams contributing to $ G^{\rm 3pt}_{\mathcal{O}_{\Gamma'};\mathcal{O}_{\Gamma\Tilde{\Gamma}};\mathcal{O}_{\Gamma''}} ((\vec{z},z_4),(\vec{z'},z_4')) \equiv \langle {\cal O}_{\Gamma'} (z) \, {\cal O}_{\Gamma \Tilde{\Gamma}} (0) {\cal O}_{\Gamma''} (z') \rangle$, up to order $\mathcal{O}(g^2)$, $({\cal O}_{\Gamma} (z) \equiv \bar \psi_{f_i}(z) \Gamma \psi_{f_j}(z))$.}
\label{fig4Q2Q2Q:3pt}
\end{center}
\end{figure}

We calculate the Green's functions perturbatively; the multiloop diagrams for two-point and three-point Green’s functions contributing up to order ${\cal O}(g^2)$ are shown in Figs.~\ref{fig4Q4Q:2pt} and \ref{fig4Q2Q2Q:3pt}, respectively.

We apply a variant of GIRS, summing over time slices of operator-insertion points for better statistical results:
\bea
\label{2pt}
\Tilde{G}^{\rm 2pt}_{\mathcal{O}_{\Gamma \Tilde{\Gamma}};\mathcal{O}_{\Gamma' \Tilde{\Gamma'}}} (z_4) &\equiv& \int d^3 \vec{z} \ G^{\rm 2pt}_{\mathcal{O}_{\Gamma \Tilde{\Gamma}};\mathcal{O}_{\Gamma' \Tilde{\Gamma'}}} (\vec{z},z_4), \\
\Tilde{G}^{\rm 3pt}_{\mathcal{O}_{\Gamma'}; \mathcal{O}_{\Gamma\Tilde{\Gamma}}; \mathcal{O}_{\Gamma''}} (z_4, z_4') &\equiv& \int d^3 \vec{z} \int d^3 \vec{z'} \ G^{\rm 3pt}_{\mathcal{O}_{\Gamma'};\mathcal{O}_{\Gamma\Tilde{\Gamma}};\mathcal{O}_{\Gamma''}} ((\vec{z},z_4),(\vec{z'},z_4')).
\label{3pt}
\eea

In this work, we present specific choices that involve all two-point Green's functions and select three-point Green's functions which lead to the smallest mixing contributions (off-diagonal matrix elements of \( Z^{S=\pm 1}_{lm} \) and \( \mathcal{Z}^{S=\pm 1} \)).

In the case of the Parity Violating operators ($\mathcal{Q}_i$), the $5 \times 5$ mixing matrix is block diagonal for both $S=+1$ and $S=-1$. This means that we only need to calculate 9 elements in total for each $S$, as the mixing occurs within smaller subsets of operators. Specifically, there are three subsets of operators that mix: $\{\mathcal{Q}_1\}$ alone, $\{\mathcal{Q}_2, \mathcal{Q}_3\}$, and $\{\mathcal{Q}_4, \mathcal{Q}_5\}$. The first subset, $\{\mathcal{Q}_1\}$, contains only one operator, which is thus multiplicatively renormalizable. Therefore, only one condition is required, and this can be obtained from two-point Green's functions. The second and third subsets each contain two operators, so four conditions are required for each subset to determine the mixing coefficients. Out of these four conditions, only three can be extracted from two-point Green's functions, while the remaining condition will be derived from three-point Green's functions.
These conditions are defined as follows [$t \equiv z_4, t' \equiv z_4^\prime$, cf. Eqs.~(\ref{2pt}) - (\ref{3pt})]:
{\small{
\bea
 [\Tilde{G}^{\rm 2pt}_{\mathcal{Q}_1^{S=\pm 1}; \mathcal{Q}_1^{S=\pm 1}} (t)]^{\rm GIRS} &\equiv& [(\mathcal{Z}_{11}^{S \pm 1})^{\rm GIRS}]^2 \ \Tilde{G}^{\rm 2pt}_{\mathcal{Q}_1^{S=\pm 1}; \mathcal{Q}_1^{S=\pm 1}} (t) = [\Tilde{G}^{\rm 2pt}_{\mathcal{Q}_1^{S=\pm 1}; \mathcal{Q}_1^{S=\pm 1}} (t)]^{\rm tree},  \\
 {[\Tilde{G}^{\rm 2pt}_{\mathcal{Q}_i^{S=\pm 1}; \mathcal{Q}_j^{S=\pm 1}} (t)]}^{\rm GIRS} &\equiv& \sum_{k,l=2}^3 \ (\mathcal{Z}_{ik}^{S \pm 1})^{\rm GIRS} (\mathcal{Z}_{jl}^{S \pm 1})^{\rm GIRS} \ \Tilde{G}^{\rm 2pt}_{\mathcal{Q}_k^{S=\pm 1}; \mathcal{Q}_l^{S=\pm 1}} (t) = [\Tilde{G}^{\rm 2pt}_{\mathcal{Q}_i^{S=\pm 1}; \mathcal{Q}_j^{S=\pm 1}} (t)]^{\rm tree}, \\
{[\Tilde{G}^{\rm 2pt}_{\mathcal{Q}_i^{S=\pm 1}; \mathcal{Q}_j^{S=\pm 1}} (t)]}^{\rm GIRS} &\equiv& \sum_{k,l = 4}^5 \ (\mathcal{Z}_{ik}^{S \pm 1})^{\rm GIRS} (\mathcal{Z}_{jl}^{S \pm 1})^{\rm GIRS} \ \Tilde{G}^{\rm 2pt}_{\mathcal{Q}_k^{S=\pm 1}; \mathcal{Q}_l^{S=\pm 1}} (t) = [\Tilde{G}^{\rm 2pt}_{\mathcal{Q}_i^{S=\pm 1}; \mathcal{Q}_j^{S=\pm 1}} (t)]^{\rm tree}. 
\eea}}
The two conditions involving three-point Green's functions are defined as:
\bea
[\Tilde{G}^{\rm 3pt}_{\mathcal{O}_\Gamma;\mathcal{Q}_i^{S=\pm 1}; \mathcal{O}_{\Gamma \gamma_5}} (t,t')]^{\rm GIRS} &\equiv& Z_{\mathcal{O}_\Gamma}^{\rm GIRS} \ Z_{\mathcal{O}_\Gamma \gamma_5}^{\rm GIRS} \ \sum_{k=2}^3 \ (\mathcal{Z}_{ik}^{S \pm 1})^{\rm GIRS} \ \Tilde{G}^{\rm 3pt}_{\mathcal{O}_\Gamma; \mathcal{Q}_k^{S=\pm 1}; \mathcal{O}_{\Gamma \gamma_5}} (t,t') \nonumber \\
&=& [\Tilde{G}^{\rm 3pt}_{\mathcal{O}_\Gamma;\mathcal{Q}_i^{S=\pm 1}; \mathcal{O}_{\Gamma \gamma_5}} (t,t')]^{\rm tree} \quad (i = \text{2 or 3}), \\
{[\Tilde{G}^{\rm 3pt}_{\mathcal{O}_\Gamma;\mathcal{Q}_i^{S=\pm 1}; \mathcal{O}_{\Gamma \gamma_5}} (t,t')]}^{\rm GIRS} &\equiv& Z_{\mathcal{O}_\Gamma}^{\rm GIRS} \ Z_{\mathcal{O}_\Gamma \gamma_5}^{\rm GIRS} \ \sum_{k=4}^5 \ (\mathcal{Z}_{ik}^{S \pm 1})^{\rm GIRS} \ \Tilde{G}^{\rm 3pt}_{\mathcal{O}_\Gamma; \mathcal{Q}_k^{S=\pm 1}; \mathcal{O}_{\Gamma \gamma_5}} (t,t') \nonumber \\
&=& [\Tilde{G}^{\rm 3pt}_{\mathcal{O}_\Gamma;\mathcal{Q}_i^{S=\pm 1}; \mathcal{O}_{\Gamma \gamma_5}} (t,t')]^{\rm tree}  \quad (i = \text{4 or 5}).
\eea
In three-point Green's functions, $t$ and $t'$ denote the GIRS renormalization scales, and it is often convenient to set $t' = t$ to reduce the complexity of the conditions.
The option that gives the smallest off-diagonal coefficients includes the following renormalized three-point functions:
\begin{equation*}
\Tilde{G}^{\rm 3pt}_{S;\mathcal{Q}_2^{S=\pm 1}; P} (t,t), \qquad \Tilde{G}^{\rm 3pt}_{S;\mathcal{Q}_5^{S=\pm 1}; P} (t,t),
\end{equation*}

For the Parity Conserving operators ($Q_i^{S=\pm 1}$), the $5 \times 5$ mixing matrix is not block diagonal, and thus, we need to calculate all 25 elements for both $S=+1$ and $S=-1$. This requires 25 conditions for each case, which can be extracted from two-point and three-point Green’s functions. From the two-point Green’s functions alone, we can extract 15 conditions. The remaining 10 conditions are obtained from three-point Green’s functions. Therefore, the full set of renormalization conditions is constructed by combining both two-point and three-point Green's function calculations.
The two-point Green's function conditions are given by:
{\small{
\bea
[\Tilde{G}^{\rm 2pt}_{Q_i^{S=\pm 1}; Q_j^{S=\pm 1}} (t)]^{\rm GIRS} &\equiv& \sum_{k,l = 1}^5 (Z_{ik}^{S \pm 1})^{\rm GIRS} (Z_{jl}^{S \pm 1})^{\rm GIRS} \ \Tilde{G}^{\rm 2pt}_{Q_k^{S=\pm 1}; Q_l^{S=\pm 1}} (t) = [\Tilde{G}^{\rm 2pt}_{Q_i^{S=\pm 1};Q_j^{S=\pm 1}} (t)]^{\rm tree},
\label{cond2pt}
\eea
}}
where $i, j$ range from 1 to 5. These conditions amount to 15 equations for the 25 unknown diagonal and off-diagonal elements of the mixing matrix.

To complete the remaining conditions, we must calculate 10 three-point Green's functions, which we select from the following family:
\bea
[\Tilde{G}^{\rm 3pt}_{\mathcal{O}_\Gamma;Q_i^{S=\pm 1}; \mathcal{O}_\Gamma} (t,t')]^{\rm GIRS} &\equiv& (Z_{\mathcal{O}_\Gamma}^{\rm GIRS})^2 \ \sum_{k=1}^5 \ (Z_{ik}^{S \pm 1})^{\rm GIRS} \ \Tilde{G}^{\rm 3pt}_{\mathcal{O}_\Gamma; Q_k^{S=\pm 1}; \mathcal{O}_\Gamma} (t,t') \nonumber \\
&=& [\Tilde{G}^{\rm 3pt}_{\mathcal{O}_\Gamma;Q_i^{S=\pm 1}; \mathcal{O}_\Gamma} (t,t')]^{\rm tree},
\label{cond3pconserving}
\eea where the renormalization factors $Z_{\mathcal{O}_\Gamma}^{\rm GIRS}$ are known from previous work~\cite{Costa:2021iyv}. The chosen  three-point Green’s functions are:
\begin{eqnarray*}
&&\hspace{-1cm}  \Tilde{G}^{\rm 3pt}_{S;Q_1^{S=\pm 1}; S} (t,t), \qquad \Tilde{G}^{\rm 3pt}_{P;Q_1^{S=\pm 1}; P} (t,t), \qquad \Tilde{G}^{\rm 3pt}_{V_i;Q_1^{S=\pm 1}; V_i} (t,t), \qquad \Tilde{G}^{\rm 3pt}_{S;Q_2^{S=\pm 1}; S} (t,t), \qquad \Tilde{G}^{\rm 3pt}_{P;Q_2^{S=\pm 1}; P} (t,t), \\
&&\hspace{-1cm}  \Tilde{G}^{\rm 3pt}_{S;Q_3^{S=\pm 1}; S} (t,t), \qquad \Tilde{G}^{\rm 3pt}_{S;Q_5^{S=\pm 1}; S} (t,t), \qquad \Tilde{G}^{\rm 3pt}_{P;Q_5^{S=\pm 1}; P} (t,t), \qquad \Tilde{G}^{\rm 3pt}_{V_i;Q_5^{S=\pm 1}; V_i} (t,t), \qquad \Tilde{G}^{\rm 3pt}_{A_i;Q_5^{S=\pm 1}; A_i} (t,t).
\end{eqnarray*}

The conversion matrix $ (C_{ij}^{S \pm 1})^{\MSbar, {\rm GIRS}}  =  (Z^{S=\pm 1})^{\rm GIRS} \left[ (Z^{S=\pm 1})^{\MSbar} \right]^{-1}$, takes the form:
\begin{equation}
(C_{ij}^{S \pm 1})^{\MSbar, {\rm GIRS}} = \delta_{ij} + \frac{g_\MSbar^2}{16 \pi^2} \sum_{k=-1}^{+1} \left[ g_{ij;k}^\pm + \left(\ln (\bar{\mu}^2 t^2) + 2 \gamma_E \right) h_{ij;k}^\pm \right] N_c^k + \mathcal{O} (g_\MSbar^4),
\end{equation}
where $g_{ij;k}^\pm$ and $h_{ij;k}^\pm$ are the numerical coefficients corresponding to the conversion matrix, and $t$ is the renormalization scale in GIRS. Table~\ref{tab:CmatrixCons} lists these coefficients.
\begin{table}[ht!]
\centering
\begin{tabular}{c|c|c|c|c|c|c|c}
\hline
$i$ & $j$ & $g_{ij;-1}^{\pm}$ & $g_{ij;0}^{\pm}$ & $g_{ij;+1}^{\pm}$ & $h_{ij;-1}^{\pm}$ & $h_{ij;0}^{\pm}$ & $h_{ij;+1}^{\pm}$ \\
\hline
$1$ & $1$ & $-869/140$ & $\pm 379/140$ & $7/2$ & $3$ & $\mp 3$ & $0$ \\
$1$ & $2$ & $2$ & $\mp (723/280 - 6 \ln(2))$ & $-2$ & $0$ & $0$ & $0$ \\
$1$ & $3$ & $-723/140 + 12 \ln(2)$ & $0$ & $0$ & $0$ & $0$ & $0$ \\
$1$ & $4$ & $-4$ & $\pm 4$ & $0$ & $0$ & $0$ & $0$ \\
$1$ & $5$ & $-2$ & $\pm 2$ & $0$ & $0$ & $0$ & $0$ \\
$2$ & $1$ & $397/280 + 6 \ln(2)$ & $\pm (163/280 -6 \ln(2))$ & $-2$ & $0$ & $0$ & $0$ \\
$2$ & $2$ & $-9/2$ & $\pm 2$ & $7/2$ & $-3$ & $0$ & $0$ \\
$2$ & $3$ & $4$ & $\mp 2$ & $0$ & $0$ & $\mp 6$ & $0$ \\
$2$ & $4$ & $4$ & $\pm 8$ & $0$ & $0$ & $0$ & $0$ \\
$2$ & $5$ & $-2$ & $0$ & $0$ & $0$ & $0$ & $0$ \\
$3$ & $1$ & $-1$ & $\pm 1$ & $0$ & $0$ & $0$ & $0$ \\
$3$ & $2$ & $1$ & $\pm 99/280$ & $0$ & $0$ & $0$ & $0$ \\
$3$ & $3$ & $-38/35$ & $\pm 2$ & $251/140$ & $-3$ & $0$ & $3$ \\
$3$ & $4$ & $4$ & $\pm 239/280$ & $-321/140$ & $0$ & $0$ & $0$ \\
$3$ & $5$ & $0$ & $\mp 239/560$ & $0$ & $0$ & $0$ & $0$ \\
$4$ & $1$ & $-1$ & $\pm 1$ & $0$ & $0$ & $0$ & $0$ \\
$4$ & $2$ & $1$ & $\mp 239/280$ & $0$ & $0$ & $0$ & $0$ \\
$4$ & $3$ & $4$ & $\pm 2$ & $-799/140$ & $0$ & $0$ & $0$ \\
$4$ & $4$ & $-307/112 + 3 \ln(2)$ & $\pm 169/140$ & $251/140$ & $-3$ & $\mp 3$ & $3$ \\
$4$ & $5$ & $-269/480 + 1/2 \ln(2)$ & $\pm (869/1680 -\ln(2))$ & $0$ & $1$ & $\mp 1/2$ & $0$ \\
$5$ & $1$ & $-6$ & $\pm 6$ & $0$ & $0$ & $0$ & $0$ \\
$5$ & $2$ & $-6$ & $0$ & $0$ & $0$ & $0$ & $0$ \\
$5$ & $3$ & $0$ & $\mp 12$ & $0$ & $0$ & $0$ & $0$ \\
$5$ & $4$ & $-269/40 + 6 \ln(2)$ & $\mp (29/140 - 12 \ln(2))$ & $0$ & $12$ & $\pm 6$ & $0$ \\
$5$ & $5$ & $-1229/240 - 3 \ln(2)$ & $\pm 309/140$ & $1709/420$ & $1$ & $\mp 3$ & $-1$ \\
\hline
\end{tabular}
\caption{Numerical values of the coefficients $g_{ij;k}^{\pm}$, $h_{ij;k}^{\pm}$ appearing in $ (C_{ij}^{S \pm 1})^{\MSbar, {\rm GIRS}}$.}
\label{tab:CmatrixCons}
\end{table}

In the case of the parity-violating operators, the $5\times5$ mixing matrix is decomposed into smaller $1\times1$ and $2\times2$ blocks. The conversion matrix $ (\Tilde{C}_{ij}^{S \pm 1})^{\MSbar, {\rm GIRS}} $ takes the form:
\begin{equation}
(\Tilde{C}_{ij}^{S \pm 1})^{\MSbar, {\rm GIRS}} = \delta_{ij} + \frac{g_\MSbar^2}{16 \pi^2} \sum_{k=-1}^{+1} \left[ \tilde{g}_{ij;k}^\pm + \left(\ln (\bar{\mu}^2 t^2) + 2 \gamma_E \right) \tilde{h}_{ij;k}^\pm \right] N_c^k + \mathcal{O} (g_\MSbar^4),
\end{equation}
where $\tilde{g}_{ij;k}^{\pm}$ and $\tilde{h}_{ij;k}^{\pm}$ are the numerical coefficients corresponding to the conversion matrix. Table~\ref{tab:CmatrixViol} lists these coefficients.

\begin{table}[ht!]
\centering
\begin{tabular}{c|c|c|c|c|c|c|c}
\hline
$i$ & $j$ & $\tilde{g}_{ij;-1}^{\pm}$ & $\tilde{g}_{ij;0}^{\pm}$ & $\tilde{g}_{ij;+1}^{\pm}$ & $\tilde{h}_{ij;-1}^{\pm}$ & $\tilde{h}_{ij;0}^{\pm}$ & $\tilde{h}_{ij;+1}^{\pm}$ \\
\hline
$1$ & $1$ & $-869/140$ & $\pm 379/140$ & $7/2$ & $3$ & $\mp 3$ & $0$ \\
$1$ & $2$ & $0$ & $0$ & $0$ & $0$ & $0$ & $0$ \\
$1$ & $3$ & $0$ & $0$ & $0$ & $0$ & $0$ & $0$ \\
$1$ & $4$ & $0$ & $0$ & $0$ & $0$ & $0$ & $0$ \\
$1$ & $5$ & $0$ & $0$ & $0$ & $0$ & $0$ & $0$ \\
$2$ & $1$ & $0$ & $0$ & $0$ & $0$ & $0$ & $0$ \\
$2$ & $2$ & $-9/2$ & $0$ & $7/2$ & $-3$ & $0$ & $0$ \\
$2$ & $3$ & $0$ & $\mp 2$ & $0$ & $0$ & $\mp 6$ & $0$ \\
$2$ & $4$ & $0$ & $0$ & $0$ & $0$ & $0$ & $0$ \\
$2$ & $5$ & $0$ & $0$ & $0$ & $0$ & $0$ & $0$ \\
$3$ & $1$ & $0$ & $0$ & $0$ & $0$ & $0$ & $0$ \\
$3$ & $2$ & $0$ & $\pm 99/280$ & $0$ & $0$ & $0$ & $0$ \\
$3$ & $3$ & $-38/35$ & $0$ & $251/140$ & $-3$ & $0$ & $3$ \\
$3$ & $4$ & $0$ & $0$ & $0$ & $0$ & $0$ & $0$ \\
$3$ & $5$ & $0$ & $0$ & $0$ & $0$ & $0$ & $0$ \\
$4$ & $1$ & $0$ & $0$ & $0$ & $0$ & $0$ & $0$ \\
$4$ & $2$ & $0$ & $0$ & $0$ & $0$ & $0$ & $0$ \\
$4$ & $3$ & $0$ & $0$ & $0$ & $0$ & $0$ & $0$ \\
$4$ & $4$ & $-307/112 + 3 \ln(2)$ & $\pm 169/140$ & $251/140$ & $-3$ & $\mp 3$ & $3$ \\
$4$ & $5$ & $-269/480 + 1/2 \ln(2)$ & $\pm (869/1680 -\ln(2))$ & $0$ & $1$ & $\mp 1/2$ & $0$ \\
$5$ & $1$ & $0$ & $0$ & $0$ & $0$ & $0$ & $0$ \\
$5$ & $2$ & $0$ & $0$ & $0$ & $0$ & $0$ & $0$ \\
$5$ & $3$ & $0$ & $0$ & $0$ & $0$ & $0$ & $0$ \\
$5$ & $4$ & $-269/40 + 6 \ln(2)$ & $\mp (29/140 - 12 \ln(2))$ & $0$ & $12$ & $\pm 6$ & $0$ \\
$5$ & $5$ & $-1229/240 - 3 \ln(2)$ & $\pm 309/140$ & $1709/420$ & $1$ & $\mp 3$ & $-1$ \\
\hline
\end{tabular}
\caption{Numerical values of the coefficients $\tilde{g}_{ij;k}^{\pm}$, $\tilde{h}_{ij;k}^{\pm}$ appearing in $ (\Tilde{C}_{ij}^{S \pm 1})^{\MSbar, {\rm GIRS}}$.}
\label{tab:CmatrixViol}
\end{table}

\section{Summary -- Future plans}
In this work, we studied the one-loop renormalization of four-quark operators in $\Delta F = 2$ processes using the Gauge Invariant Renormalization Scheme (GIRS). We calculated two-point Green's functions for pairs of four-quark operators and three-point Green’s functions with one four-quark and two bilinear operators, using dimensional regularization. Operator mixing was addressed through renormalization conditions on these Green's functions. Various valid renormalization prescriptions within GIRS were identified, applicable to both perturbative and nonperturbative data. We derived one-loop conversion matrices connecting nonperturbative GIRS results to $\MSbar$ and present a choice that minimizes mixing contributions. 

A natural extension of this work will involve the investigation of four-quark operators with $\Delta F = 1$ and $\Delta F = 0$. This investigation also entails the mixing with lower-dimensional operators, including extended quark bilinear operators, the chromomagnetic operator, and the energy-momentum tensor.

\begin{acknowledgments}
The project (EXCELLENCE/0421/0025) is implemented under the programme of social cohesion ``THALIA 2021-2027'', co-funded by the European Union through the Cyprus Research and Innovation Foundation (RIF). M.~Constantinou acknowledges financial support from the U.S. Department of Energy, Office of Nuclear Physics, Early Career Award under Grant No.\ DE-SC0020405. The results are generated within the FEDILA software (project: CONCEPT/0823/0052), which is also implemented under the same ``THALIA 2021-2027'' programme, co-funded by the European Union through RIF. G.S. and M.~Constantinou acknowledge funding under the project 3D-nucleon, contract number EXCELLENCE/0421/0043, co-financed by the European Regional Development Fund and the Republic of Cyprus through RIF.
\end{acknowledgments}


\begin{thebibliography}{99}

\bibitem{Buchalla:1995vs}
G.~Buchalla, A.~J.~Buras and M.~E.~Lautenbacher,
\emph{Weak decays beyond leading logarithms},
 \href{https://doi.org/10.1103/RevModPhys.68.1125}{Rev. Mod. Phys. \textbf{68} (1996) 1125} 
 [{\tt arXiv:hep-ph/9512380 [hep-ph]}].

\bibitem{USQCD:2019hyg}
C.~Lehner \textit{et al.} [USQCD],
\emph{Opportunities for Lattice QCD in Quark and Lepton Flavor Physics},
\href{https://doi.org/10.1140/epja/i2019-12891-2}{Eur. Phys. J. A \textbf{55} (2019) 195} 
[{\tt arXiv:1904.09479 [hep-lat]}].

\bibitem{FlavourLatticeAveragingGroupFLAG:2021npn}
Y.~Aoki \textit{et al.} [Flavour Lattice Averaging Group (FLAG)],
\emph{FLAG Review 2021},
\href{https://doi.org/10.1140/epjc/s10052-022-10536-1}{Eur. Phys. J. C \textbf{82} (2022) 869} 
[{\tt arXiv:2111.09849 [hep-lat]}].

\bibitem{LHCb:2022aki}
R.~Aaij \textit{et al.} [LHCb],
\emph{Observation of a Resonant Structure near the Ds+Ds- Threshold in the B+\textrightarrow{}Ds+Ds-K+ Decay},
\href{https://doi.org/10.1103/PhysRevLett.131.071901}{Phys. Rev. Lett. \textbf{131} (2023) 071901}
[{\tt arXiv:2210.15153 [hep-ex]}].
 
\bibitem{Costa:2021iyv}
M.~Costa, I.~Karpasitis, T.~Pafitis, G.~Panagopoulos, H.~Panagopoulos, A.~Skouroupathis and G.~Spanoudes,
\emph{Gauge-invariant renormalization scheme in QCD: Application to fermion bilinears and the energy-momentum tensor},
\href{https://doi.org/10.1103/PhysRevD.103.094509 }{Phys. Rev. D \textbf{103} (2021) 094509} 
[{\tt arXiv:2102.00858 [hep-lat]}].

\bibitem{Lin:2024mws}
J.~Lin, W.~Detmold and S.~Meinel,
\emph{Position-space renormalization schemes for four-quark operators in HQET},
\href{https://link.springer.com/article/10.1007/JHEP07(2024)188}{JHEP \textbf{07} (2024) 188}
[{\tt arXiv:2404.16191 [hep-lat]}].

\bibitem{Constantinou:2024wdb}
M.~Constantinou, M.~Costa, H.~Herodotou, H.~Panagopoulos and G.~Spanoudes,
\emph{Gauge-invariant renormalization of four-quark operators in lattice QCD},
\href{https://journals.aps.org/prd/abstract/10.1103/PhysRevD.110.074506}{Phys. Rev. D \textbf{110} (2024) 074506} 
[{\tt arXiv:2406.08065 [hep-lat]}].

\bibitem{Frezzotti:2004wz}
R.~Frezzotti and G.~C.~Rossi,
\emph{Chirally improving Wilson fermions. II. Four-quark operators},
\href{https://doi.org/10.1088/1126-6708/2004/10/070}{JHEP \textbf{10} (2004) 070}
{\tt{[arXiv:hep-lat/0407002 [hep-lat]]}.}

\end{thebibliography}

\end{document}